\documentclass{optica-article}

\journal{opticajournal} 

\articletype{Research Article}

\usepackage{lineno}
\usepackage{xfrac, nicefrac}
\usepackage{upgreek}
\usepackage{graphicx}
\usepackage{subcaption}

\begin{document}

\newcommand{\ail}{\mathrm{APSF_{il}}}
\newcommand{\adet}{\mathrm{APSF_{det}}}
\newcommand{\aism}{\mathrm{APSF_{ISM}}}

\newcommand{\pil}{\mathrm{PSF_{il}}}
\newcommand{\pdet}{\mathrm{PSF_{det}}}
\newcommand{\pism}{\mathrm{PSF_{ISM}}}
\newcommand{\pconf}{\mathrm{PSF_{conf}}}
\newcommand{\ppxl}{\mathrm{PSF}_{x_p}}

\newcommand{\apil}{\mathrm{(A)PSF_{il}}}
\newcommand{\apdet}{\mathrm{(A)PSF_{det}}}
\newcommand{\apism}{\mathrm{(A)PSF_{ISM}}}

\newcommand{\sapil}{\mathrm{\sigma_{\apil}}}
\newcommand{\sapdet}{\mathrm{\sigma_{\apdet}}}

\newcommand{\spil}{\mathrm{\sigma_{il}}}
\newcommand{\spdet}{\mathrm{\sigma_{det}}}


\title{Tsang's resolution enhancement method \\for imaging with focused illumination}

\author{Alexander Duplinskiy,\authormark{1,*} Jernej Frank,\authormark{1} Kaden Bearne\authormark{1} and A. I. Lvovsky\authormark{1}}

\address{\authormark{1} Department of Physics, University of Oxford, Oxford, OX1 3PU, UK}

\email{\authormark{*}al.duplinskiy@gmail.com} 


\begin{abstract*}

A widely tested approach to overcoming the diffraction limit in microscopy without disturbing the sample relies on substituting widefield sample illumination with a structured light beam. This gives rise to confocal, image-scanning and structured-illumination microscopy methods. On the other hand, as shown recently by Tsang and others, subdiffractional resolution at the detection end of the microscope can be achieved by replacing the intensity measurement in the image plane with spatial mode demultiplexing. In this work we study the combined action of Tsang's method with image scanning. We experimentally demonstrate superior lateral resolution and enhanced image quality compared to either method alone. This result paves the way for integrating spatial demultiplexing into existing microscopes, contributing to further pushing the boundaries of optical resolution.

\end{abstract*}

\section{Introduction}

The ability to achieve resolution beyond the diffraction limit of light has revolutionized the field of microscopy. Super-resolution techniques have enabled researchers to visualize structures and processes with unprecedented detail, leading to new discoveries and insights. The demand for better resolving optical systems has been traditionally driven by biological studies, where delicate nature of samples often does not allow direct interaction. Therefore, non-invasive methods operating in the far-field have become highly popular as practical means to enhance resolution \cite{liu2022super, valli2021seeing, schermelleh2019super, heintzmann2017super, wu2018faster}.

Confocal microscopy is one of the most established and widely implemented super-resolution techniques that operate through non-uniform illumination \cite{sheppard1977image, wilson1980imaging, pawley2006handbook}. In this method, the sample is scanned under a focused illumination beam while a pinhole placed in the image plane acts as a spatial filter, allowing only the light from the center of the illuminated object area to pass through. The intensity distribution recorded with a bucket detector with respect to the position of the scan forms high-resolution image of the object.

Building upon this concept, image scanning microscopy (ISM) \cite{cox1982improvement, muller2010image, GREGOR201974} replaces the pinhole and the bucket detector with a detector array or a camera \cite{cox1982improvement, sheppard1988super, muller2010image}. Snapshots recorded at each scanning step are combined into the final image using the pixel reassignment routine \cite{Chen16, mcgregor2015post}. While providing the same lateral resolution, ISM eliminates the optical loss caused by the pinhole in the confocal setup. 

An alternative resolution improvement method, proposed by Tsang {\it et al.}\cite{tsang2016quantum}, is based upon decomposing
the incoming field in the detection plane into an orthonormal basis of transverse modes, such as Hermite-Gaussian (HG) modes, and measuring the amplitude or intensity of each  mode. From these measurements, the original object can be
reconstructed \cite{yang2016far}. This method, to which we refer as Hermite-Gaussian Imaging (HGI),  emerged from fundamental discoveries in quantum measurement theory in recent years \cite{tsang2016quantum,yang2016far,tsang2017subdiffraction,tsang2018subdiffraction,tsang2019quantum,tsang2019resolving,pushkina2021superresolution,bearne2021confocal,zanforlin2022optical,ozer2022reconfigurable,Frank23}. This method allows one to not only achieve sub-Rayleigh
precision, but also, in some cases, reach the ultimate resolution limits allowed by quantum mechanics \cite{tsang2016quantum, tsang2019quantum}. 

In contrast to ISM, confocal microscopy and other methods that  achieve resolution enhancement by applying transverse structure to the illuminating beam, HGI aims at improving the opposite end of the imaging system --- the detection of light. Hence it is natural to ask whether both approaches can be combined with a cumulative effect. 
In this work we implement this combination and demonstrate that applying HGI in the scanning paradigm results in resolution and image quality improvement compared to both techniques used independently. 

\section{Image scanning for lateral resolution improvement}


The effective point-spread function (PSF) of an imaging system with a focused illumination beam can be expressed as a product of the illumination and  detection PSFs 
: $\pism = \pdet \cdot \pil$. Assuming that both PSFs are Gaussian with the widths $\sigma_{\rm{il}}$ and $\sigma_{\rm{det}}$, we have (see Appendix):

\begin{equation} \label{PSF_ISM}
\pism(x)  \propto \exp\left(-\frac{x^2}{2\sigma^2_{\rm{det}}}\right) \exp\left(-\frac{x^2}{2\sigma^2_{\rm{il}}}\right) \propto \exp\left(-\frac{x^2}{2\sigma^2_{\rm{ISM}}}\right)
\end{equation}
with
\begin{equation} \label{(A)PSF_ISM}
 \sigma_{\rm{ISM}}= \frac{\sigma_{\rm{il}}\sigma_{\rm{det}}}{\sqrt{\sigma^2_{\rm{il}}+\sigma^2_{\rm{det}}}} .
\end{equation}

Usually the same objective lens is used both to focus the beam onto the specimen and to collect the light from it, meaning equal numerical apertures (NA) in illumination and detection. If the wavelengths of illumination and detection are the same, we have $\spil = \spdet$ and  $\sigma_{\rm{ISM}} = \frac{1}{\sqrt{2}} \spdet$. Reduction of the Gaussian PSF width by a $\sqrt{2}$ factor corresponds to the same resolution enhancement in terms of the generalized Rayleigh criterion. But if one of the PSFs, either in detection or illumination, is significantly narrower than the other, then the resolution $\sigma_{\rm{ISM}}$ of ISM is close to the width of that narrower PSF. This could be the case if the  wavelength of detection is greater than that of illumination, e.g.~due to the Stokes shift in fluorescence microscopy. For example, if   $\sigma_{\rm{il}}=0.5 \sigma_{\rm{det}}$ , then  $\sigma_{\rm{ISM}}\approx 0.89\sigma_{\rm{il}}$.
For coherent imaging, the amplitude point spread function (APSF) serves as the equivalent to the PSF in incoherent scenario. It characterizes the field distribution in the image plane generated by a point source in the object plane. The resulting intensity distribution of the image can be calculated as a square of the convolution of the object field with the APSF. For confocal and image scanning cases, the expressions provided above for the incoherent scenario can be adopted by simply replacing PSFs with APSFs (see Table \ref{compare_img}). 

\begin{table}[h]
    \centering
    \setlength{\extrarowheight}{2pt}

    \begin{tabular}{>{\bfseries}  c | c | c}
        \multicolumn{1}{c}{} & \textbf{Widefield} & \textbf{Image scanning} \\
        \hline
        \textbf{Incoherent} & $\mathrm{I = Obj \circledast \pdet}$& $\mathrm{I = Obj \circledast \left( \pdet \cdot \pil \right)}$ \\
        \hline
        \textbf{Coherent} & $\rm{I = \big| obj \circledast \adet \big|^2}$  & $\mathrm{I = \big| obj \circledast \left(\adet \cdot \ail \right)\big|^2}$ \\
    \end{tabular}
    \caption{Image formation comparison for coherent and incoherent, widefield and image scanning setups. I - intensity function of the image; Obj, obj - intensity and amplitude reflections of the object, respectively.
    }
    \label{compare_img}
\end{table}

It's noteworthy to acknowledge that while the Gaussian approximation offers a convenient means for estimating resolution improvement, it is not entirely precise. The APSF of a circular aperture is a $\mathrm{Jinc}(\cdot)$ function ($\mathrm{Jinc}^2(\cdot)$ for the PSF in the incoherent case). The resolution enhancement according to the generalized Rayleigh criterion for incoherent imaging and identical illumination and detection PSFs in this case is about 1.53 
\cite{cox1982improvement, sheppard2013superresolution}.

\section{Hermite-Gaussian imaging and scanning}
Tsang's initial proposal to overcome the diffraction limit by examining the HG modes addressed the task of distinguishing two point sources in the far field and estimating the distance between them \cite{tsang2016quantum}. It was shown that if, instead of the standard intensity measurement, one isolates the first order HG mode and measures its intensity, the Fisher information per photon  remains constant even when the sources approach each other. Following this insightful discovery, extensive efforts, both theoretical  and experimental, have been invested in exploring  broader applications of this technique, extending beyond single-parameter scenarios to encompass imaging tasks. Here we briefly recap the main concepts of HGI from Ref.~\cite{yang2016far}.

\begin{figure}[htbp]
  \centering
  \includegraphics[width=0.8\textwidth]{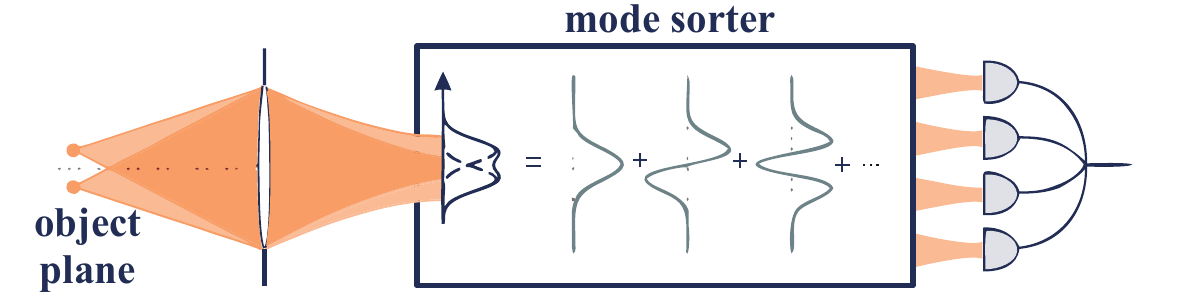}
\caption{The concept of Hermite-Gaussian imaging via spatial mode demultiplexing. The measurement in carried out by decomposing the optical field into the basis of Hermite-Gaussian modes, with amplitudes measured in the coherent case or intensities for incoherent imaging. The outcomes of the measurements can be correlated with the geometrical moments of the sample in the object plane.}
\label{fig:HGI}
\end{figure}

HGI decomposes the field in the image plane into the Hermite-Gaussian basis, defined as 
\begin{equation} \label{field_HGI}
\phi_{mn}(x, y) = \frac{1}{2^{n}n!} \frac{1}{\sqrt{2\pi\sigma^2}}H_{n}\left(\frac{x}{\sqrt{2}\sigma}\right) H_{m}\left(\frac{y}{\sqrt{2}\sigma}\right) \mathrm{e}^{-\frac{x^2 + y^2}{4\sigma_{\rm det}^2}},
\end{equation}
where $H_n(x)$ corresponds to Hermite polynomial of the order $n$. One then measures the amplitude (for coherent imaging) or intensity (for incoherent imaging) of each mode (Fig.~\ref{fig:HGI}). These measurements allow one to access the corresponding geometrical moments of the field/intensity distribution of the initial object. 
For the coherent scenario one can directly extract the object field distribution from the measured amplitudes. Incoherent HGI measurements yield only even moments of the object intensity distribution, so the HG basis needs to be augmented with HG 
 mode superpositions to access odd moments \cite{tsang2017subdiffraction}. Eventually in both cases the obtained moments are used to reconstruct the field or intensity of the object \cite{pushkina2021superresolution, Frank23}. The resolution achieved with HGI depends on the number of modes used in the experiment and is limited only by the signal-to-noise ratio as the detectable amount of light in each mode drops exponentially with the mode order. Previous experiments on HGI under widefield illumination have shown resolution improvement by factors of 2--3 compared to direct imaging (DI) both in incoherent and coherent scenarios \cite{yang2016far,pushkina2021superresolution,zanforlin2022optical,ozer2022reconfigurable,Frank23}. 

In this work, we implement HGI in the ISM setting by replacing the CCD  with HG measurements. At each scanning step, HGI reconstructs a snapshot of the object under restricted illumination  with improved resolution. Subsequent pixel reassignment allows us to form a final image of the object based on individual snapshots. 

Reconstructing each snapshot from HG measurements is a complex process, in practice requiring a neural network \cite{pushkina2021superresolution}. However, the reconstructed image can be approximated as a convolution of the object with an effective detection (A)PSF. 
Hence we can describe the resolution of HGI-based ISM using the same formalism as DI-based ISM as discussed in the previous section.  

\section{Experiment}
We use a macroscopic reflective coherent imaging setting on an optical table akin to Refs.~\cite{pushkina2021superresolution, Frank23}. For illumination, we utilise a laser source with a central wavelength of 795 nm. To approach the resolution limit we decrease the numerical aperture (NA) of the imaging system down to $0.71\cdot10^{-3}$ by placing an iris in the imaging path at a significant distance ($\sim2.5$~m) away from the object. This iris is the entrance pupil of the imaging system and leads to a direct imaging resolution of  $\sigma_{\rm det,DI} = 0.21\uplambda/\rm{NA}\simeq 0.23$ mm.

\begin{figure}[htbp]
  \centering
  \includegraphics[width=0.8\textwidth]{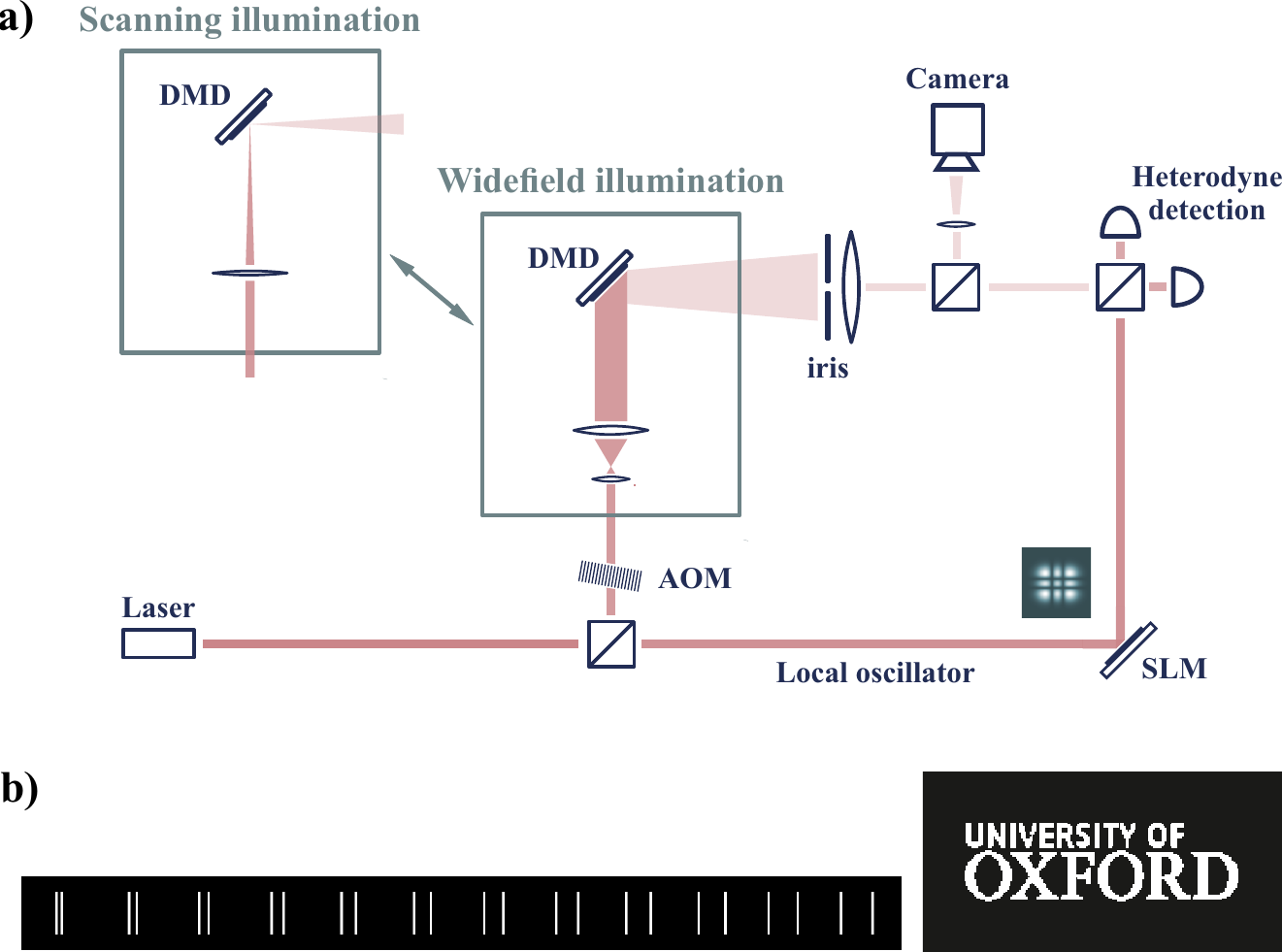}
\caption{a) Optical setup for comparing DI and HGI under widefield and focused scanning illumination. The objects to image are displayed on the DMD. Measurement in HG basis is carried out by heterodyne detector with the local oscillator sequentially prepared in different HG modes by the SLM. b) Bitmaps displayed on the DMD for resolution estimation: Oxford University logo and pairs of lines with separation varying from 20 to 130 DMD pixels. }
\label{fig:setup}
\end{figure}

A digital micromirror device (DMD) is used to display various  objects composed of binary ``logical" pixels. Each logical pixel is a square with a side of 10 physical DMD pixels (75.6 $\mu$m). The area utilized  on the DMD in all experiments is a square of size $210\times210$ physical pixels (hereafter referred to as ``frame"). We study two objects [Fig.~\ref{fig:setup}(b)]. First, to quantitatively assess the lateral resolution in the experiment, we employ a set of parallel line pairs of 175 DMD pixels lengths with the  centre-to-centre separation varying  from 20 to 130 DMD pixels in steps of  10 DMD pixels. Second, we use the Oxford University logo ($1680\times 630$ DMD pixels) as an example of a more complex object. This object is displayed on the DMD as a sequence of frames as described below.

Unlike typical confocal or ISM setups, the illumination beam and the light reflected from the sample do not share the same optical path in our experiment. This grants us flexibility to switch between different illumination settings independently of the detection scheme. We compare HGI and DI both with widefield and scanning approaches. Experiments with each of the four combinations are described below.


\subsection{Widefield DI}
DI is performed by imaging the DMD onto the CCD camera with a set of lenses. The collimated illumination beam is larger than the DMD screen and can be considered uniform within the displayed frame region.  For the line separation measurement, each pair is displayed and acquired separately. To image  the university logo, which is larger than the DMD frame size, we cut it into a number of overlapping frames and display them one by one. After collecting the images, we crop out the central part before stitching them together. Removing the edges in each picture minimizes the impact of optical aberrations on the image quality and allows for fairer comparison of the methods purely from the resolution prospective \cite{pushkina2021superresolution}.

\subsection{Scanning DI}
Our DI scanning experiment mimics the ISM setting on a macroscopic scale. We focus the illumination beam so that the waist of $\ail$ on the DMD is significantly less then the displayed frame size. To emulate the real-world scenario when  $\ail$ and $\adet$ are equal, we independently measure the corresponding intensity distributions and match them. 
The measured $\sigma$-parameter of $\left|\adet\right|^2$ and $\left|\ail\right|^2$ is 28 DMD pixels (212 $\mu$m) and is slightly less than the value estimated from NA (30 DMD pixels).


Once the illumination is set in the centre of the DMD frame, it stays constant and scanning is emulated by displaying displaced objects on the DMD. At each step, the pattern on the DMD is shifted by 20 DMD pixels and the snapshot is captured by the camera. After collecting all individual snapshots we recombine them using the pixel reassignment routine (as described in the Appendix) to get an improved-resolution image of the initial object.

\subsection{Widefield HGI}
For widefield HGI we reproduce  the coherent imaging experiment described in the previous work \cite{pushkina2021superresolution}. Measurement in the HG basis is carried out by means of heterodyne detection, with the local oscillator (LO) sequentially prepared in each of the 441 HG modes of orders 0 to 20 in both transverse dimensions. 

The light from the laser is split into two paths. One of them is used to illuminate the sample whilst the other serves as the LO. Acousto-optic modulator (AOM) introduces a frequency shift of 80 MHz to the illumination beam and a spatial light modulator (SLM) is utilized for the LO mode preparation. After reflecting from the DMD and passing through the iris, the optical signal is recombined with the LO mode on a beamsplitter and the output is measured with a balanced detector to extract the amplitudes and phases corresponding to each HG mode for every individual sample.

To compensate for various imperfections of the optical setup, we use a fully connected neural network (NN) with four hidden layers that transforms HG measurements into images. The training set comprises $2\cdot10^4$ patterns of random patterns, lines and ellipses displayed on the DMD. The corresponding photocurrents measured for different HG modes are the inputs of the NN. The labels are the ground truth images that are slightly smeared to avoid overfitting  \cite{pushkina2021superresolution}. Once trained, the NN is utilized to reconstruct the objects of interest which were not part of the training set.

\subsection{Scanning HGI}
The optical setup for scanning HGI is modified in a similar manner as for the scanning DI counterpart by narrowing the light beam illuminating the DMD. The NN training set is the same as for the widefield HGI, however for label generation we take into account narrow illumination of the samples, so the label images fade to black outside the illuminated region. 

Once the NN is trained, the test set of objects is imaged. Each object is displayed with multiple shifts along both axes to implement scanning. For each step, the snapshot is reconstructed using the NN. Subsequently, individual snapshots are combined through pixel reassignment to derive the final image.

\section{Results and discussion}
To estimate the resolution value for each particular method, we utilize the set of line pairs and interpolate the pixel value to find the distance for the generalised Rayleigh limit. 
The estimated resolution for DI is $111\pm1$ DMD pixel being slightly better than the coherent Rayleigh limit estimation of $0.84\uplambda/\rm{NA}\simeq 120$ DMD pixels based on the NA measurement. Narrowing the illumination beam and implementing scanning with pixel reassignment enhances the widefield resolution by a factor of $1.6$ ($69\pm1$ DMD pixels) as expected for coherent ISM. The HGI experiment with widefield illumination managed to achieve $46\pm1$ DMD pixel resolution, meaning a factor of 2.4 resolution improvement compared to widefield DI. Finally, HGI with scanning reached $44\pm1$ DMD pixels. The expected resolution improvement of scanning HGI compared to widefield HGI evaluated via Eq.~\eqref{PSF_ISM} with $\sigma_{\rm il}/\sigma_{\rm det}=2.4$ cannot exceed 8\%, consistent with these observations.   


\begin{figure}[htbp]
  \centering
  \includegraphics[width=\textwidth]{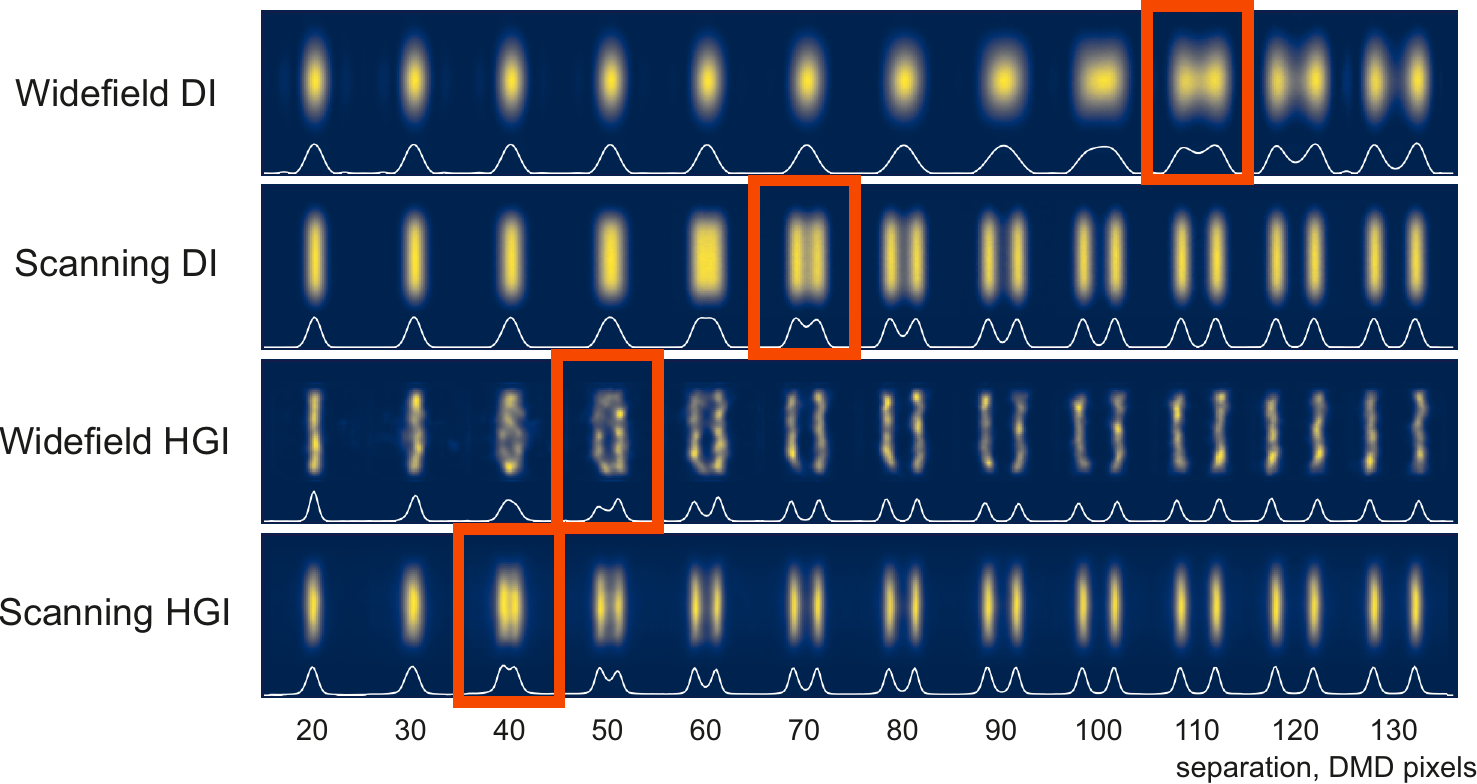}
\caption{Imaging the line pairs with gradually changing separation using DI and HGI under widefield and focused scanning illumination. All images of pairs are normalised. White line illustrates the one-dimensional profile. Red frames indicate the pairs closest to the Rayleigh limit for each method.}
\label{fig:lines}
\end{figure}

\begin{figure}[htbp]
  \centering
  \includegraphics[width=\textwidth]{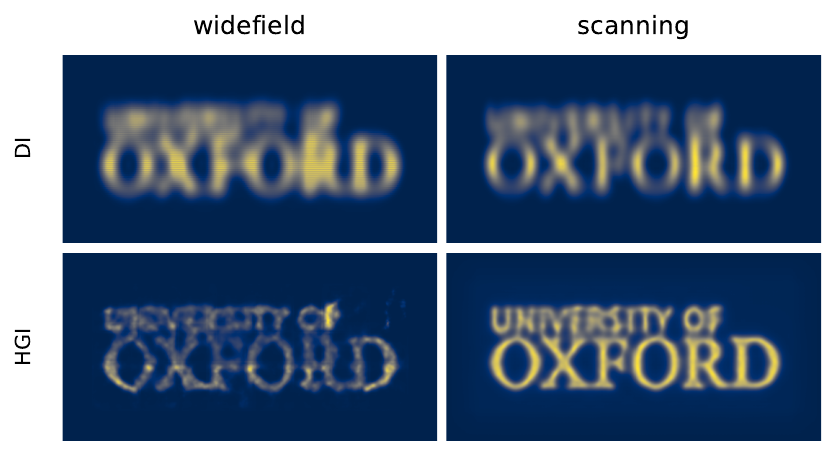}
\caption{Imaging the Oxford University logo using DI and HGI under widefield and focused scanning illumination.}
\label{fig:logo}
\end{figure}

The Oxford University logo images produced with different methods (Fig. \ref{fig:logo}) follow the same trend as the line pairs, with scanning HGI achieving the best quality. Notably, both the line pairs and the logo reveal artifacts in widefield HGI that are distinct from conventional optical blur.

As discussed, the more resolution improvement we get from HGI the less ISM can add to it. 
However, as one can observe, the scanning significantly reduces the impact of HGI artifacts on the final image and consequently improves its quality. The reason is that, as mentioned, experimental image formation in HGI can only approximately be described as a convolution with a Gaussian PSF. In practice, the HGI point source response might depend on various factors such as the position in the object plane and the presence of other sources. For example, as seen in Fig.~\ref{fig:lines}, HGI seems to achieve better results in the centre of the picture than around the edges. Augmenting HGI with scanning provides translational invariance and averages out the artifacts produced by each individual reconstructed snapshot.


Since HGI incorporates the NN to reconstruct images, it is challenging to disentangle the improvements achieved by the measurement itself from the potential effects of the NN. To isolate the NN's contribution, we applied a similar NN to the snapshots acquired in the DI experiments. The network failed to produce any noticeable enhancement to the resolution both for widefield and scanning data. This aligns with our previous unsuccessful efforts of employing the conventional Richardson-Lucy deconvolution algorithm for images acquired in the same coherent experiment \cite{pushkina2021superresolution}. We thus conclude that  the observed resolution enhancement is due to the new physics of HGI rather than the effects of NN.



\section{Conclusion}
In this study, we introduced a novel approach by integrating the HGI technique with the conventional ISM superresolution method. We experimentally demonstrate how both methods benefit from complementing each other. Although in the HGI setting the additional resolution enhancement due to ISM is limited, we observe substantial improvement in the overall image quality. The ultimate resolution improvement for coherent imaging in our experiment reached a factor of 2.5. This result makes the combined strategy promising for imaging of real-world specimens. 

\section*{Acknowledgements}
The project is funded by BBSRC grant BB/X004317/1. JF is supported by the European Union’s Horizon 2020 research and innovation programme under the Marie Skłodowska-Curie grant agreement
No 956071. KB is supported by the Clarendon Fund scholarship.

\bibliography{sample}

\section*{Appendix. Image scanning microscopy}
\label{App}
Let us consider the incoherent case. We assume the illumination beam stays centered  in the object plane, so its intensity distribution is ($\pil(x)$) and the sample, whose spatial distribution is denoted by $\mathrm{Obj}(\cdot)$ is displaced by $S$. The resulting intensity distribution in the image plane can be expressed as
\begin{equation} \label{general_ISM}
 I(y|S) = \int \mathrm{Obj}(x+S) \cdot \pil(x) \cdot \pdet(y-x)   dx.
\end{equation}


Substituting the illumination and detection PSF profiles we obtain
\begin{equation}
\pil(x) \cdot \pdet(y-x)\propto  \exp\left(-\frac{x^2}{2\sigma^2_{\rm{il}}}\right)  \cdot \exp\left(-\frac{(x-y)^2}{2\sigma^2_{\rm{det}}}\right)\propto   \exp\left(-\frac{y^2}{2\sigma^2_{\rm{det}}}\right)\cdot \mathrm{exp}\left(-\frac{\left(x-\frac{\sigma^2_{\mathrm{ISM}}}{\sigma^2_{\mathrm{det}}}y\right)^2}{2\sigma^2_{\rm{ISM}}}\right),
\end{equation}
where \begin{equation} 
 \sigma_{\rm{ISM}}= \frac{\sigma_{\rm{il}}}{\sqrt{\sigma^2_{\rm{il}}+\sigma^2_{\rm{det}}}} \cdot \sigma_{\rm{det}}.
\end{equation}
and we used the proportionality sign to contain all constants (independent of $x$, $y$, and $S$). Hence
\begin{equation} 
 I(y|S) \propto \exp\left(-\frac{y^2}{2\sigma^2_{\rm{det}}}\right)\int \mathrm{O
 bj}(x+S) \cdot \mathrm{exp}\left(-\frac{\left(x-\frac{\sigma^2_{\mathrm{ISM}}}{\sigma^2_{\mathrm{det}}}y\right)^2}{2\sigma^2_{\rm{ISM}}}\right)  dx.
\end{equation}
Let us rescale the acquired single-shot image by defining a new function:
\begin{equation} 
I' (y | S ) = I \left(\frac{\sigma^2_{\mathrm{det}}}{\sigma^2_{\mathrm{ISM}}}y \middle| S\right).
\end{equation}
This rescaled image is a convolution of the object at the current position with a narrower Gaussian PSF. The result of the convolution is also multiplied by a decaying exponent meaning that the image is only produced for the illuminated part of the object:

\begin{equation} 
I' (y | S )  \propto \exp\left(-\frac{y^2}{2\sigma^2_{\rm{det}}}\right)\int \mathrm{Obj}(x+S) \cdot \mathrm{exp}\left(-\frac{\left(x-y\right)^2}{2\sigma^2_{\rm{ISM}}}\right)  dx.
\end{equation}
We can replace the integration variable for $x' = x+S$ :
\begin{equation} 
I' (y | S )  \propto \exp\left(-\frac{y^2}{2\sigma^2_{\rm{det}}}\right)\int \mathrm{Obj}(x') \cdot \mathrm{exp}\left(-\frac{\left(x'-y-S\right)^2}{2\sigma^2_{\rm{ISM}}}\right)  dx'.
\end{equation}
To obtain the full image, we sum all the single-shot images, displaced by the corresponding shift $S$:
\begin{equation} 
I_{\rm ISM}(y)= \int I'(y-S | S ) dS \propto \iint\exp\left(-\frac{(y-S)^2}{2\sigma^2_{\rm{det}}}\right) \mathrm{Obj}(x') \cdot \mathrm{exp}\left(-\frac{\left(x'-y\right)^2}{2\sigma^2_{\rm{ISM}}}\right)  dx'dS.
\end{equation}
We notice that the double integral can be separated:
\begin{equation} 
I_{\rm ISM}(y)\propto \int\exp\left(-\frac{(y-S)^2}{2\sigma^2_{\rm{det}}}\right) dS \cdot \int \mathrm{Obj}(x') \mathrm{exp}\left(-\frac{\left(x'-y\right)^2}{2\sigma^2_{\rm{ISM}}}\right)  dx'.
\end{equation}



The first integral is a Gaussian integral: its value is constant and can be absorbed into the proportionality sign:
\begin{equation} \label{sum_over_s}
I_{\mathrm{ISM}}(y') = \sum_{i} I'\left(y'\middle|S_i\right) \propto\int \mathrm{Obj}(x') \cdot \mathrm{exp}\left(-\frac{\left(x'-y'\right)^2}{2\sigma^2_{\rm{ISM}}}\right)  dx'.
\end{equation}
We see that the result of this procedure,  known as pixel reassignment, is a convolution of the object intensity distribution with $\pism(\cdot)$.

\end{document}